\documentclass[12pt]{article}

\usepackage{amsmath,amssymb,graphicx} 
\usepackage{pstricks}
\usepackage{cite}

\newcommand{\beq}{\begin{eqnarray}}
\newcommand{\eeq}{\end{eqnarray}}

\newcommand{\centeron}[2]{{\setbox0=\hbox{#1}\setbox1=\hbox{#2}\ifdim
\wd1>\wd0\kern.5\wd1\kern-.5\wd0\fi \copy0
\kern-.5\wd0\kern-.5\wd1\copy1\ifdim\wd0>\wd1
                                  \kern.5\wd0\kern-.5\wd1\fi}}
\newcommand{\ltap}{\>\centeron{\raise.35ex\hbox{$<$}}
                          {\lower.65ex\hbox{$\sim$}}\>}
\newcommand{\gtap}{\>\centeron{\raise.35ex\hbox{$>$}}
                          {\lower.65ex\hbox{$\sim$}}\>}

\newcommand\ZZ{\hbox{\zfont Z\kern-.4emZ}}
\font\zfont = cmss10 

\newcommand{\eref}[1]{eq.\ (\ref{e.#1})}

\newcommand{\cref}[1]{Chapter \ref{c.#1}}

\def\nn{\nonumber \\}

\newcommand{\bnl}{\right . & \nonumber \\ & \left .}

\def\beq{\begin{equation}}
\def\eeq{\end{equation}}
\newcommand{\ba}{\begin{array}}
\newcommand{\ea}{\end{array}}
\newcommand{\bea}{\begin{eqnarray}}
\newcommand{\eea}{\end{eqnarray} }
\newcommand{\bal}{\begin{align}}
\newcommand{\eal}{\end{align}}
\def\bi{\begin{itemize}}
\def\ei{\end{itemize}}
\def\ben{\begin{enumerate}}
\def\een{\end{enumerate}}
\def\beq{\begin{equation}}
\def\eeq{\end{equation}}
\def\bc{\begin{center}}
\def\ec{\end{center}}
\def\bt{\begin{table}}
\def\et{\end{table}}
\def\btb{\begin{tabular}}
\def\etb{\end{tabular}}
\newcommand{\bvec}{\left ( \ba{c}}
\newcommand{\evec}{\ea \right )}

\def\cl{{\mathcal L}}

\def\cm{{\mathcal M}}

\def\tev{\, {\rm TeV}}

\def\mass2{mass${}^2$}

\def\pa{\partial}

\newcommand{\tr}{\mathrm T \mathrm r}

\newcommand{\ti}{\tilde}
\def\hc{{\rm h.c.}}

\def\ov{\overline}

\begin{document}
\begin{titlepage}

\vskip2.5cm
\begin{center}
{\huge \bf A Simple Flavor Protection for RS} \vspace*{0.1cm}
\end{center}
\vskip0.2cm

\begin{center}
{\bf Csaba Cs\'aki$^{a}$, Adam Falkowski$^{b,c}$, and Andreas
Weiler$^{a}$}

\end{center}
\vskip 8pt

\begin{center}
$^{a}$ {\it Institute for High Energy Phenomenology\\
Newman Laboratory of Elementary Particle Physics\\
Cornell University, Ithaca, NY 14853, USA } \\
\vspace*{0.3cm}

$^b$ { \it CERN Theory Division, CH-1211 Geneva 23, Switzerland} \\
\vspace*{0.3cm} $^c$ {\it Institute of Theoretical Physics, Warsaw
University,
          \\ Ho\.za 69, 00-681 Warsaw, Poland }

\vspace*{0.3cm}

{\tt  csaki@lepp.cornell.edu, adam.falkowski@cern.ch, weiler@lepp.cornell.edu}
\end{center}

\vglue 0.3truecm

\begin{abstract}
\vskip 3pt \noindent We present a simple variation of warped
flavor models where the hierarchies of fermion masses and mixings
are still explained but dangerous flavor violating effects in
the Kaon sector are greatly reduced. The key new ingredients are two
horizontal U(1) symmetries. These symmetries forbid flavor violation
in the down quark sector (with the exception of small IR brane
localized kinetic mixing terms for the left-handed quarks) while
allowing for flavor violation in the up quark sector. The leading
flavor constraints come from $D^0 - \bar D^0$ mixing, and are
safely satisfied for the KK mass scale of order 3 TeV. 
Our analysis of the flavor constraints also includes the constraints due to the (usually
ignored) localized kinetic mixing terms. 
We also comment on the effects of the additional U(1) gauge bosons. 

\end{abstract}

\end{titlepage}

The Randall-Sundrum (RS) model \cite{RS} provides an interesting
framework for new physics beyond the Standard Model (SM). Not only
does it address the electroweak hierarchy problem because the Higgs
sector localized on the IR brane has effectively a TeV cut-off
scale, but it also offers an explanation of the SM fermion mass
hierarchies via localized profiles of fermions in the extra
dimension \cite{GN,GP}. As SM fermions get their masses from Yukawa
interactions with the Higgs on the IR brane, one can localize the
light fermions close to the UV brane to make their effective Yukawa
couplings hierarchically small.
The simplest ``anarchic approach" assumes that the Yukawa couplings  in the 5D theory are all ${\cal
O}(1)$ in natural units.
Then the localization of the SM fermions leads to the effective 4D mass matrices which naturally incorporate hierarchies both in the mass eigenvalues and the CKM mixing angles \cite{H}.

Theories of flavor quite generally predict new sources of flavor
violation. In the RS model, flavor changing neutral currents
(FCNC's) arise already at the tree level, due to flavor violating
couplings of the Kaluza-Klein (KK) modes to the quark mass
eigenstates. These same KK modes play the important role of
stabilizing the  electroweak scale, and naturalness implies that the
lightest KK mode should not be heavier than $\sim 3$ TeV. At first
sight this spells disaster for the model, since tight experimental
bounds on FCNCs would normally imply a much higher KK scale,
$M_{KK}> 10^3$--$10^4$ TeV. However, the RS model has a built-in
flavor protection usually referred to as RS-GIM \cite{GP,APS}: the
localization mechanism responsible for the mass hierarchies also
suppresses FCNC's. RS-GIM is powerful enough to suppress below the
experimental sensitivity almost all effective $\Delta F=2$
four-fermion operators generated by the exchange of KK modes. The
exception is the left-right (LR) operator $\mathcal{O}_4=(\bar{s}_L
d_R) (\bar{s}_R d_L)$ whose imaginary part is tightly bounded by
measurements of CP violation in the kaon sector. It turns out that
the coefficient of this operator comes out roughly a factor of 100
too large for a $\sim 3$ TeV KK gluon \cite{CFW}, if one assumes no
tuning among the input parameters. Thus, some form of flavor
symmetry is  required if we insist that the RS model addresses both
the electroweak naturalness problem and the flavor hierarchies.

One possible  approach is to require a full-fledged 5D GIM
mechanism, where all tree-level FCNC's are completely absent. This
is indeed possible \cite{CCGMTW} using $U(3)^3$ flavor symmetries in
the bulk, and assuming that the only source of flavor violation are
fermion kinetic terms localized on the UV brane. The price is,
however, that the hierarchy of fermion masses and mixings is left
completely unexplained. One would prefer a solution that
incorporates just enough flavor symmetries to eliminate the
dangerous sources of flavor violation, while still allowing for the
generation of flavor hierarchies. An attempt in this direction is
the ``5D MFV" proposal of \cite{FPRa}, which postulates that only
two spurions are responsible for breaking the flavor symmetry,
generating both the bulk mass matrices and the brane Yukawa
matrices. This assumption does not suppress FCNCs by itself and an
additional alignment of bulk masses and brane Yukawa terms has to be
imposed~\cite{FPRa,inprep}. Recently, an economical model based on a
$U(3)$ symmetry acting in the bulk and broken on the IR brane was
proposed in \cite{S}. This symmetry, while allowing for down quark
mixing via Yukawa couplings on the IR brane, enforces a degeneracy
of the bulk masses in the down right quark sector which ensures
universal couplings to vector KK modes.  Flavor
symmetries have also proved successful in warped models of the
lepton sector, where they can explain the observed neutrino mixing
patterns and charged lepton flavor hierarchies without excessive
lepton flavor violation \cite{leptons}.

In this paper we propose a set of symmetries for the quark sector
that strongly protects the down quarks from flavor violations  both
in the right- and left-handed down sector, even after taking into
account the effect of brane localized kinetic mixing terms. The idea
is to use a horizontal $U(1)$ symmetry acting in the down quark
sector. This symmetry acts in the bulk and on the IR brane, while it
is broken on the UV brane. It aligns the bulk masses with the brane
Yukawa matrices in the down sector: both are required to be diagonal
in the basis in which the $U(1)$ charges are diagonal. This idea is
similar to the SUSY alignment models~\cite{NS}, with the aim of
moving all (or most) of flavor violation into the up sector. There
is an immediate problem with this requirement: since the left-handed
doublet quarks are charged under this $U(1)$, it seems we would need
to assign charges in the right-handed up sector too to allow for the
up-type Yukawa couplings. This would indeed be the case in the
simplest RS-type model where only one bulk multiplet per generation
hosts a quark doublet, and our  $U(1)$ would forbid {\em all} flavor
mixing and force the CKM matrix to be the unit matrix. However, one
can construct well motivated models in which each SM quark doublet
is embedded in several bulk multiplets. This is e.g. the case in the
model  of ref. \cite{CDP} where the extended structure is introduced
in order to ensure the custodial protection of the $Z \ov b_L b_L$
vertex \cite{ACDP}. In such extended models we are able to implement
a horizontal $U(1)$ with desired properties. For reasons that will
become clear in a moment we will need in fact two separate $U(1)$
symmetries, one of which is broken on the UV brane, and the other on
the IR brane.

\renewcommand{\arraystretch}{1.5}
\begin{table}
   \begin{center}
\begin{tabular}{ c c c c c}
              & $\Psi_u$ & $\Psi_{q_u}$ & $\Psi_{q_d}$ & $\Psi_{d}$  \\ \hline
 $U(1)_q \,(+,-)$ &  $\cdot$     &  $ {q}_i$        & $q_i$  &  $\cdot$ \\
 $U(1)_d \,(-,+)$ &  0     &  0              & $d_i$          &  $d_i$
\end{tabular}
\end{center}
\caption{ Charge assignments under the horizontal $U(1)$ symmetries.}
\end{table}
\renewcommand{\arraystretch}{1}

The model we propose is inspired by ref. \cite{CDP}, though here we deal with a smaller gauge symmetry group in the bulk and therefore we can  utilize a more economical set of bulk fields.
We consider 5D gauge theory in a slice of AdS$_5$. We parametrize the space-time by
the conformal coordinates
\begin{equation}
ds^2=\left( \frac{R}{z}\right)^2 (dx_\mu dx_\nu \eta^{\mu\nu} -dz^2) \, ,
\end{equation}
where the AdS curvature is $R$,  and the coordinate $z$ of the extra
dimension  runs between $R<z<R'$, $z=R$ corresponding to the UV
(Planck) brane and $z=R'$ to the IR (TeV) brane. $R'/R \sim 10^{16}$
is the large number  that sets the hierarchy between the Planck and
the TeV scale. Our gauge group in the bulk is $SU(3)_c \times
SU(2)_L\times SU(2)_R \times U(1)_X$ broken by boundary conditions
on the UV brane down to $SU(3)_c \times SU(2)_L\times U(1)_Y$
\cite{ADMS}. The Higgs field  $\Phi$ transforming as $(1,2,2)_0$ is
localized on the IR brane.
Each generation of SM quarks is embedded in  four 5D quark multiplets. The doublets are embedded in two
bi-fundamental  (under $SU(2)_L\times SU(2)_R$) quarks $\Psi_{q_u,q_d}$.
This realization incorporates the custodial protection of the $Z \ov b_L b_L$ vertex \cite{ACDP} if the couplings of the two SU(2)'s are equal.
The SM singlet quarks are embedded into $SU(2)_L \times SU(2)_R$ singlets $\Psi_{u,d}$.
The representation under the full gauge group and the boundary
conditions are chosen as
\begin{equation}
{\bf (3,2,2)_{2/3}:} \ \Psi_{q_u} = \bvec q_{u} [\pm,+] \quad \ti
q_{u}[-,+]   \evec \qquad {\bf (3,2,2)_{-1/3}:}\ \Psi_{q_d} =  \bvec
\ti q_{d}[-,+] \quad q_{d}[\pm,+] \evec \nonumber\end{equation}
\begin{equation} {\bf (3,1,1)_{2/3}:} \ \Psi_u =  \bvec
u^c[-,-]  \evec \qquad {\bf (3,1,1)_{-1/3}:} \ \Psi_d =  \bvec
d^c[-,-]  \evec
\end{equation}
In the square bracket we indicated the boundary conditions on the UV and IR branes, where $[+/-]$ denotes the
right/left chirality of the bulk fermion vanishing on the brane.
$[\pm]$ stands for mixed boundary conditions for the $SU(2)_L$
doublets:
\begin{equation}
\theta q_{u,L}(0) -   q_{d,L}(0)  = 0 \qquad
q_{u,R}(0) + \theta^\dagger q_{d,R}(0)  = 0
\end{equation}
where $\theta$ is a $3\times 3$ matrix.
These boundary conditions could be obtained by coupling the combination $\theta q_{u,L} - q_{d,L}$
to three UV localized  right-handed doublets $\psi_R$ via the mass term
$\Lambda_0 \ov \psi_R (\theta q_{u,L} - q_{d,L})$ and then taking the limit of $\Lambda_0 \to \infty$.
The effect of these boundary conditions is that the two bulk fields $q_{u,d}$ host only
one left-handed zero mode $q_L(x)$. There are also right-handed zero
modes $u_R(x)$, $d_R(x)$ hosted by $u^c$, $d^c$. After electroweak
symmetry breaking these zero modes obtain masses from the IR brane
localized Yukawa interactions
\begin{equation}
\cl_{yuk} = -   (R^4/R'{}^3)
\delta(z-R') \left ( \tr\left [\ov \Psi_{q_u,L} \Phi \right] \ti Y_u
\Psi_{u,R} + \tr\left [\ov \Psi_{q_d,L}  \Phi \right] \ti Y_d
\Psi_{d,R} \right ) + \hc
\end{equation}
In the absence of flavor symmetries the Yukawa matrices are expected
to be anarchical. We keep this assumption for the up-type Yukawa
matrix, but we impose non-trivial structure on the down-type Yukawa
matrix. We achieve this by introducing a $U(1)_d$ symmetry acting on
the $X = -1/3$ quarks with generation dependent charges
$(d_1,d_2,d_3)$, the same for $\Psi_{q_d}$ and for  $\Psi_{d}$.
There are no constraints on the values of the charges other than
$d_i \neq d_j$ for $i \neq j$. The symmetry is valid in the bulk and
on the IR brane but must be broken on the UV brane  to allow for
non-zero $\theta$. This symmetry will ensure that $c_{q_d}, c_d$ and
the down Yukawa coupling $\tilde{Y}_d$ are simultaneously diagonal,
and also forbids off-diagonal kinetic terms on the IR brane
involving $\psi_{q_d}$ and $\psi_d$.  The $U(1)_d$ can be global or
local. In the latter case there is a corresponding gauge boson with
$(-+)$ boundary conditions.
We will later comment on the phenomenology of such a gauge boson.

In this model there is another  source of flavor violation in the
down sector if the $\theta$ matrix that sets the UV boundary
conditions is non-diagonal. To protect us from that we need to
introduce another $U(1)_q$ symmetry that acts on the UV brane and in
the bulk, but it has to be  broken on the IR brane in order to allow
non-diagonal up-type Yukawa couplings. Under this symmetry,
$\Psi_{q_{u}}$ and $\Psi_{q_{d}}$ should have equal charges in each
generation, but the charges of different generations should be
different, $q_i \neq q_j$ for $i \neq j$. We can also charge the
rest of the multiplets, which will have the effect of  forbidding
all off-diagonal brane kinetic terms on the UV brane. This symmetry
will then ensure that $c_{q_u},c_{q_d}$ and $\theta $ are all
simultaneously diagonal. Again, the symmetry can be global or local,
in the latter case the $U(1)_q$ gauge boson has $(+-)$ boundary
conditions.

The two U(1) symmetries together ensure that  $c_{q_u},c_{q_d}, c_u,
c_d, \tilde{Y}_d$ and $\theta$ are all diagonal while allowing for
off-diagonal components in $\tilde{Y}_u$. Moreover, off-diagonal
kinetic terms  are not allowed on the UV brane (if we also charge
$\psi_{u,d}$ under U(1)$_q$). On the IR brane, the symmetries still
allow for two off-diagonal brane kinetic terms:
\begin{equation} \label{e.bkt}
\cl = \frac{R^4}{{R'}^3}  i \delta \left ( \ov \Psi_{q_u} \ti K_q
\pa {\!\!\! /}\Psi_{q_u} + \ov \Psi_{u} \ti K_u \pa {\!\!\!
/}\Psi_{u} \right )
\delta(z - R)
\end{equation}
We cannot decree that these terms vanish because they are generated
by loop effects, but it is natural to assume that the coefficients
are loop suppressed.
It is important to note, that these kinetic mixing terms are present in every RS flavor model, but they are ignored because they are usually sub-leading. However, if the leading sources are suppressed this might turn out to be the 
dominant contribution. In particular in models with partial flavor protection \cite{FPRa,S} these operators will be the sources of the leading flavor constraints.
In the following we assume $\ti K \sim 1$ and the
NDA value of the dimensionless coefficient
\begin{equation}
\delta \sim Y_*^2
\frac{\Lambda R'}{16 \pi^2},
\end{equation}
where $Y_*$ is the typical amplitude of
the entries in $\ti Y_{u,d}$, and $\Lambda$ is the cutoff scale for
the Yukawa interaction, usually assumed to be around the second KK
mode. The off-diagonal terms in $\ti K_{{q_u},u}$ lead to the
kinetic mixing among the up-type quarks. This is a subleading effect
compared to the mass mixing generated by the Yukawa terms so that it
can be safely neglected. However, $\ti K_{q_u}$ leads also to the
kinetic mixing among the left-handed down quarks. This is a source
of flavor violation in the down quark sector which we need to take
into account. The improvement with respect to the standard RS flavor
scenario is that the brane kinetic terms affect mostly the
left-handed down quarks (so that there are no large LR currents) and
are loop suppressed.

We are ready to analyze the flavor structure of our model.
Let us write down the mass matrix for the SM quark fields $u_{L,R},
d_{L,R}$ which are zero modes of the bulk fermions. The localization
of zero modes depends on the fermion bulk mass customarily written
in terms of dimensionless $c$-parameters:
\begin{equation}
\frac{R^4}{z^5} \sum_{x = c_{q_u},c_{q_d},c_u,c_d}  \bar{\Psi}_{x} c_x \Psi_{x} .
\end{equation}
where each $c$ is a $3 \times 3$ matrix in the flavor space, $c_x = {\rm diag}(c_x^1,c_x^2,c_x^3)$.
The zero modes are embedded into the 5D fields as
\bea
\label{e.dmmap} 
q_{u,L}(x,y) &\to&    R'{}^{-1/2}  \left( \frac{z}{R}\right)^{2}  \left( \frac{z}{R'}\right)^{-c_{q_u}}  f_{q_u}
C_\theta  q_L(x)
\nn
q_{d,L}(x,y) &\to&    R'{}^{-1/2}  \left( \frac{z}{R}\right)^{2}  \left( \frac{z}{R'}\right)^{-c_{q_d}} f_{q_d}
S_\theta q_L(x)
\nn
u_R^{c}(x,y) &\to&   R'{}^{-1/2}  \left( \frac{z}{R}\right)^{2}  \left( \frac{z}{R'}\right)^{c_u} f_{-u} u_R(x)
\nn
d_R^{c}(x,y) &\to&   R'{}^{-1/2}  \left( \frac{z}{R}\right)^{2}  \left( \frac{z}{R'}\right)^{c_d} f_{-d} d_R(x)
\eea
where $f_x = {\rm diag}(f(c_{x^1}),f(c_{x^2}),f(c_{x^3}))$ and $f(c)$ is defined as
\begin{equation}
\label{e.rsff}
f(c)=\frac{\sqrt{1-2 c}}{\left[1-(\frac{R'}{R})^{2c-1}\right]^{\frac{1}{2}}}.
\end{equation}
Furthermore, we traded $\theta$ for the diagonal matrices $C_\theta$,
$S_\theta$ defined as
\begin{equation}
C_\theta = {1 \over \sqrt{1 + |\hat\theta|^2}}
\qquad
S_\theta = {\hat \theta \over \sqrt{1 + |\hat\theta|^2}}
\qquad
\hat \theta = \theta {f_{q_u} \over  f_{q_d}} (R'/R)^{c_{q_u} - c_{q_d}},
\end{equation}
which decide how the doublet zero modes are distributed between
the two bulk multiplets. In this basis the kinetic terms for the
zero-mode quarks are not diagonal due to the brane kinetic terms in
\eref{bkt}. We can safely neglect the kinetic mixing of the up
quarks since the mixing via the Yukawa couplings is much larger, but
we should take into account the kinetic mixing of the down quarks
which is the leading effect. The kinetic terms for the left-handed
quark are given by,
\begin{equation}
K = 1  + \delta \, C_\theta f_{q_u} \ti K_q f_{q_u} C_\theta .
\end{equation}
To diagonalize them, we need to perform the Hermitian transformation
$d_L(x) \to H d_L(x)$ where $H K H = 1$.\footnote{The hermitian
matrix $H$ can be decomposed as $H=U^\dagger_{d_L} N U_{d_L}$, where
$U_{d_L}$ is the unitary matrix diagonalizing the kinetic mixing
term $K$ and $N$ is a positive  diagonal matrix rescaling the
kinetic terms to be canonical.} After that, we can plug the zero
mode profiles into the brane Yukawas and replacing the Higgs with
its vev to obtain the SM mass matrices,
\bea \label{e.smmm}
\cm_u^{SM} &=& {v \over \sqrt{2}} C_\theta f_{q_u} \ti Y_u f_{-u}
\nn \cm_d^{SM} &=& {v \over \sqrt{2}} H S_\theta f_{q_d} \ti Y_d
f_{-d}
\eea

To proceed, we need to find the bi-unitary rotations that diagonalize these mass matrices.
We start with the up quark mass matrix.
The natural expectation is that $\ti Y_u$ is anarchical, and below we work under
this assumption.
On the other hand, as there are no non-abelian flavor symmetries
to enforce the equality of the c-parameters, $f_{q_u,q_d,-u,-d}$ are expected to be hierarchical.
Below we assume  $f_{q_{u}^1} \ll f_{q_{u}^2} \ll
f_{q_{u}^3}$ and   $f_{-u^1} \ll f_{-u^2} \ll f_{-u^3}$. Then the up
quark matrix can be diagonalized as
$\cm_u^{SM} = L^u (m_u)_{\rm diag} R^u{}^\dagger$,  where the elements of
the rotation matrices are approximately given by
\begin{equation}
|L_{ij}^u| \sim \frac{f_{q_{u}^i}}{f_{q_{u}^j}}, \ \
| R_{ij}^u| \sim \frac{f_{-u^i}}{f_{-u^j}}, \ \ {\rm} \ i\leq j.
\end{equation}
Since the dominant source of flavor mixing comes from the up sector, the CKM matrix is approximately given by the  hermitian conjugate of the rotation
matrix $L^u$.
Thus the hierarchy in the CKM matrix elements is  set by the $c_{q_u}$
parameters and we need to choose them such that
\begin{equation}
f_{q_u^2}/f_{q_u^3} \sim \lambda^2, \qquad f_{q_u^1}/f_{q_u^3} \sim \lambda^3,
\end{equation}
where $\lambda \sim \sin \theta_{C} \sim 0.2$. The remaining $c$
parameters have to be adjusted to  reproduce the quark masses. In
this model we have actually more $c$ parameters than quark masses to
fit so several options are possible. For the purpose of
illustration, we propose here one particular pattern. We assume that
all $c$ parameters for the third generation are such that the zero
modes are localized in the IR, that is $c_{q_{u,d}}^3 < 1/2$,
$c_{u,d}^3
-1/2$, whereas the first two generations are localized in UV:
$c_{q_{u,d}}^{1,2} > 1/2$, $c_{u,d}^{1,2} < -1/2$. At this point,
important observables depend on the relation between  $c_{q_u}$ and
$c_{q_d}$. We  define three dimensionless numbers
\begin{equation} \alpha_i =
(R'/R)^{c_{q_u}^i - c_{q_d}^i}
\end{equation}
As we will see in the moment, in this scenario the $m_b/m_t$ ratio is set by $|\alpha_3|$. The small
ratio can be obtained by choosing $c_{q_u}^3 <  c_{q_d}^3$ so as to
make $|\alpha_3| \ll 1$, rather then localizing the right bottom in UV
as in the standard approach to RS flavor.
This implies that $\hat \theta_3 \ll 1$, whereas $\hat \theta^{1,2} \sim 1$, as the ratio of
$f's$ and the exponent cancel in the later case.
For the moment we leave the options open for $\alpha_{1,2} \sim f_{q_d^{1,2}}/f_{q_u^{1,2}}$: they
can be smaller or larger than 1.

Next, we discuss the down quark mass matrix.
Although the down-type Yukawa couplings are  diagonal, the mass matrix is slightly tipped by the Hermitian transformation $H$.
The off-diagonal terms in the  kinetic matrix are suppressed  by both a loop factor and by RS-GIM and,
in consequence, the Hermitian rotation matrix is close to the unit matrix
$(H)_{ij} \sim \delta_{ij} - {\delta \over 2}  f_{q_u^i} f_{q_u^j}$.
The unitary rotations $L^d$ and $R^d$ that diagonalize down mass matrix  have smaller off-diagonal terms than the
corresponding up quark rotations,
\begin{equation}
|L_{ij}^d| \sim    {\delta \over 2} f_{q_u^i} f_{q_u^j}
\qquad
| R_{ij}^d| \sim  \delta {m_d^i \over m_d^j} f_{q_u^i} f_{q_u^j} \ \  \ i <  j.
\end{equation}

The eigenvalues of the quark mass matrices are of the order of
\begin{equation}
(m_{u})_{\rm diag} \sim  {v Y_* \over \sqrt 2}   f_{q_{u}} f_{-u}
\qquad (m_{d})_{\rm diag} \sim  {v Y_* \over \sqrt 2}  \alpha
f_{q_{u}} f_{-d}.
\end{equation}
Thus, the bottom to top quark mass ratio is ${m_b/m_t}  \sim
\alpha_3 {f_{-d^3} /f_{-u^3}} $ which is of order $\alpha_3$ as long
as $f_{-d^3} \sim f_{-u^3} \sim 1$.
In order to reproduce the up quark masses, the ratios of $f_{-u}$'s
should be adjusted as \begin{equation} {f_{-u^1} \over f_{-u^3}} \sim {m_u \over
m_t} \lambda^{-3} \qquad {f_{-u^2} \over f_{-u^3}} \sim {m_c \over
m_t} \lambda^{-2}. \end{equation} This fixes the magnitude of  the rotation
angles in  the right-handed up quark sector. As for the remaining
down quark masses, we need \begin{equation}
\alpha_1 {f_{-d^1} \over f_{-u^3}} \sim {m_d \over m_t} \lambda^{-3}
\qquad
\alpha_2 {f_{-d^2} \over f_{-u^3}} \sim {m_s \over m_t} \lambda^{-2},
\end{equation}
we can adjust $\alpha_{1,2}$ or $c_d^{1,2}$ or both to match the experimental values.

We move to discussing the FCNCs in our model. Typically, the
strongest constraints come from $\Delta F = 2$  processes mediated
by tree-level exchange of the KK gluons, as their coupling to matter
fields is usually stronger than that of other vector KK modes.
Thus the first task is to determine the off-diagonal couplings of the KK gluon to the
zero-mode quarks. In the original flavor basis the coupling depends
on the parameter $c$ of the zero mode, and is approximately given by
\cite{APS,CFW}
\begin{eqnarray}
g_{x} \approx  
g_{s*} \left(
-\frac{1}{\log R'/R} +  f_x^2 \,\gamma(c_x) + \delta f_x \ti K_x f_x
\right).
\label{GBcoupling}
\end{eqnarray}
where $g_{s*}$ is the bulk $SU(3)$ gauge couplings
($g_{s*} \sim 6$ in the absence of brane kinetic terms for the $SU(3)$ gauge fields),
and $\gamma(c)$  is of order
one but it varies by order 1 values with $c$ varied in the range of
interest.
The last term is due to the IR brane kinetic terms for the fermions.
Non-diagonal couplings arise if $\ti K_x$ is non-diagonal,
or they may be generated by the unitary transformation from the flavor basis to the mass-eigenstate basis if $f_x$ is non-universal.
In our model both of these sources are present.


The leading off-diagonal couplings to the left- and right-handed up quarks can be read from
\begin{equation}
(g_{u_L}) \sim g_{s*}  L_u{}^{\dagger} f_{q_u}^2 (1 + \alpha^2 ) L_u
\qquad
(g_{u_R}) \sim g_{s*} R_u{}^{\dagger} f_{-u}^2 R_u.
\end{equation}
Performing the rotation, we find that the off-diagonal couplings of
the KK gluon to right handed quarks are given by $c_u$,
$(g_{R})_{ij} \sim f_{-u^i} f_{-u^j}$, which in turn depend on the
observable CKM mixing angles and mass ratios and on $f_{-u^3} \sim
1$. The RS-GIM mechanism is at work in the sense that the couplings
to the light quarks are suppressed by the hierarchically small
$f_{-u}$. In particular, for the up and the charm quarks,
\begin{equation}
(g_{u_R})_{12} \sim  g_{s*} {m_u  m_c \over m_t^2}  {f_{-u^3}^2 \over \lambda^{5}}
\sim  g_{s*} {2 m_u  m_c \over Y_*^2 v^2} { 1 \over \lambda^{5} f_{q_u^3}^2} ,
\end{equation}
which is suppressed by $m_u m_c/m_t^2\lambda^5 \sim 5 \cdot
10^{-5}$. For the couplings of the left-handed quarks we have some
more freedom. If all $|\alpha_i| \ll 1$ then $(g_{u_L})_{ij} \sim
g_{s*} f_{q_u^i} f_{q_u^j}$ and the couplings can be expressed in
terms of the CKM hierarchy. For $|\alpha_{1,2}| > 1$ the couplings
may be enhanced. For the up and the charm quarks we find
\begin{equation}
\label{e.dupa1}
(g_{u_L})_{12} \sim g_{s*} f_{q_u^1} f_{q_u^2}
\left (1 + \alpha_2^2 + \alpha_3^2 + {f_{q_u^1}^2 \over f_{q_u^2}^2} \alpha_1^2 \right )
\sim g_{s*} \lambda^5 f_{q_u^3}^2 \left (1 + \alpha_2^2\right )
\end{equation}
which is RS-GIM suppressed by $\lambda^5 \sim 6 \cdot 10^{-4}$.
In the second step we used our assumptions $|\alpha_3| \ll 1$, $f_{q_d^1} \ll f_{q_d^2}$.
Writing $1 + \alpha^2$ we really mean the larger of the two numbers,
since we are estimating a sum of complex contributions with random phases.

For the left-handed down quark, the off-diagonal KK gluon couplings
appear  due to the brane-kinetic terms even before diagonalizing the
kinetic terms and the mass matrix. After performing the Hermitian
rotation, the couplings become
\begin{equation} (g_{d_L}) \sim g_{s*}  H \left ( -
{1 \over \log (R'/R)}  +f_{q_u}^2 (1 + \alpha^2 ) + \delta f_{q_u}
\ti K_q f_{q_u} \right ) H
\end{equation}
One can see that the matrix  $H$, designed to  diagonalize the
kinetic terms, does not simultaneously diagonalize the KK gluon
couplings. Instead, we are left with the off-diagonal terms
$(g_L^d)_{ij} \sim  g_{s*} \delta f_{q_u^i} f_{q_u^j}$. The
subsequent unitary rotation $L_d$ does not generate any larger
terms. For the down and strange quarks we find
\begin{equation}
(g_{d_L})_{12} \sim  g_{s*} \delta \lambda^5 f_{q_u^3}^2.
\end{equation}
Besides the RS-GIM suppression, this coupling enjoys the  loop
suppression factor $\delta$. For the right-handed down quarks, the
unitary rotation $R_d$ is the only source of flavor off-diagonal
couplings who for this reason become even more  suppressed:
$(g_{d_R})_{ij} \sim g_{s*} \delta (m_d^i/m_d^j) f_{q_u^i} f_{q_u^j}
f_{-d^j}^2$, $i < j$. For the down and strange quarks
\begin{equation}
(g_{d_R})_{12} \sim g_{s*} \delta  {m_d \over m_s} \lambda^5 f_{q_u^3}^2 f_{-d^2}^2
\sim
g_{s*} \delta  {2 m_d m_s \over Y_*^2 v^2} {\lambda \over \alpha_2^2}.
\end{equation}

Having obtained the flavor violating couplings,
we can estimate the magnitude of the flavor-violating four-fermion operators
generated by a tree-level exchange of the KK gluons.
After applying the Fierz identities the effective Hamiltonian is of the form~\cite{CFW}
\begin{eqnarray}
\mathcal{H} = & \frac{1}{M_G^2} \left[
\frac{1}{6} (g_{q_L})_{ij} (g_{q_L})_{kl}
(\bar{q}_L^{i \alpha} \gamma_\mu q_{L \alpha}^j)\ (  \bar{q}_L^{k \beta} \gamma^\mu q_{L \beta}^l)
+ \frac{1}{6} (g_{q_R})_{ij} (g_{q_R})_{kl}
(\bar{q}_R^{i \alpha} \gamma_\mu q_{R \alpha}^j)\ (  \bar{q}_R^{k \beta} \gamma^\mu q_{R \beta}^l)
\bnl
- (g_{q_R})_{ij} (g_{q_L})_{kl} \left( (\bar{q}_R^{i \alpha}  q_{L\alpha}^k)\
(\bar{q}_L^{l \beta} q_{R \beta}^j)
-\frac{1}{3}  (\bar{q}_R^{i \alpha} q_{L\beta}^l)\
(\bar{q}_L^{k \beta} q_{R \beta}^j)\right)\right]
\end{eqnarray}
where $\alpha,\beta$ are color indices.
This parameterization of flavor violating new physics contributions in
terms Wilson coefficients $C(\mu)$:
\begin{eqnarray}
\mathcal{H}  &=& C^1(\mu) (\bar{q}_L^{i \alpha} \gamma_\mu q_{L
\alpha}^j)\ (  \bar{q}_L^{k \beta} \gamma^\mu q_{L \beta}^l) + \ti
C^1(\mu) (\bar{q}_R^{i \alpha} \gamma_\mu q_{R \alpha}^j)\ (
\bar{q}_R^{k \beta} \gamma^\mu q_{R \beta}^l) \nn & + &
C^4(\mu)
(\bar{q}_R^{i \alpha} q_{L  \alpha}^k)\ (\bar{q}_L^{l \beta} q_{R
\beta}^j) +   C^5(\mu) (\bar{q}_R^{i \alpha} q_{L \beta}^l)\
(\bar{q}_L^{k \beta} q_{R \alpha}^j) \nonumber
\end{eqnarray}
is subject to constraints from low-energy experiments \cite{UTfit}.
In our model the strongest constraints come
from $\Delta F = 2$ transitions between the up and the charm quarks in $D^0 - \bar D^0$ mixing.
The corresponding Wilson coefficients are given by
\begin{equation}
\label{e.dmwc} 
C_D^1(M_G) \sim {1 \over M_G^2} {g_{s*}^2 \over 6} \lambda^{10}
f_{q_u^3}^4 (1 + \alpha_2^2)^2 \qquad
\ti C_D^1(M_G) 
\sim {1 \over M_G^2} {g_{s*}^2 \over Y_*^4}  {2 m_u^2 m_c^2 \over 3 v^4} {1 \over \lambda^{10} f_{q_u^3}^{4}}
\end{equation}
\begin{equation}
C_D^4(M_G) = - 3 C_D^5(M_G) 
\sim {1 \over M_G^2} {g_{s*}^2 \over Y_*^2} {2 m_u m_c \over v^2}(1
+ \alpha_2^2)
\end{equation}
We also give the Wilson coefficients in the kaon sector
\begin{equation}
\label{e.kmwc} 
C_K^1(M_G) \sim {1 \over M_G^2}  {g_{s*}^2 \over 6} \delta^2  \lambda^{10}
f_{q_u^3}^4 \qquad \ti C_K^1(M_G) \sim {1 \over M_G^2} {g_{s*}^2 \over Y_*^4}
\delta^2 \frac{2 m_d^2 m_s^2 }{3 v^4}
{\lambda^{2} \over \alpha_2^4}
\end{equation}
\begin{equation}
C_K^4(M_G) = - 3 C_K^5(M_G) \sim {1 \over M_G^2}  {g_{s*}^2 \over Y_*^2} \delta^2
\frac{2 m_d m_s}{v^2} \frac{\lambda^{6} f_{q_u^3}^2}{\alpha_2^2}
\end{equation}

\renewcommand{\arraystretch}{1.5}
\begin{table}
\begin{center}
\begin{tabular}{@{}lllll}
\hline\hline
Parameter & Suppression & $f_{q_u^3} = 0.3$ & $f_{q_u^3} = 1$  & Bound (TeV) \\
\hline
$|C_{D}^{1}|$  & ${\sqrt{6}  \over g_{s*} \lambda^5 f_{q_u^3}^2}M_G$ &
$7.8 \cdot 10^{3} M_G$  &  $0.7 \cdot 10^{3} M_G$ & $1.2 \cdot 10^{3}$
\\
$|\ti C_{D}^{1}|$  & ${\sqrt{3} Y_*^2 v^2 \lambda^5  f_{q_u^3}^2 \over  \sqrt{2} g_{s*}  m_u m_c } M_G$ &
$1.2 \cdot 10^{3} M_G$ &$1.3 \cdot 10^{5} M_G$
& $1.2 \cdot 10^{3}$
 \\
$|C_{D}^{4}|$  & ${v Y_* \over g_{s*} \sqrt{2\, m_u m_c}}M_G$&
$ 1.2 \cdot 10^{3} M_G$ & $ 1.2 \cdot 10^{3} M_G$
& $3.5 \cdot 10^{3}$
\\
\hline
$|C_{K}^{1}|$  & $ {\sqrt{6}  \over g_{s*} \lambda^5 f_{q_u^3}^2 \delta } M_G$ &
$3.0 \cdot 10^{6} M_G$  &  $2.7 \cdot 10^{5} M_G$
& $1.5 \cdot 10^{4}$
\\
$|\ti C_{K}^{1}|$  & $ {\sqrt{3} Y_*^2 v^2  \over \sqrt{2} g_{s*} m_d m_s \lambda  \delta } M_G$ &
$1.5 \cdot 10^{10} M_G$  &  $1.5 \cdot 10^{10} M_G$
& $1.5 \cdot 10^{4}$
\\
$|C_{K}^{4}|$  & $ \frac{Y_* v}{ g_{s*} \sqrt{2 m_d m_s} \lambda^3 f_{q_u^3} \delta} M_G$ &
$2.8 \cdot 10^{7} M_G$ & $8.5 \cdot 10^{6} M_G$
& $1.6 \cdot 10^{5}$
\\ \hline
\end{tabular}
\end{center}
\caption {Flavor violation in the D- and K-meson sector.
In the second column we give the suppression scale of the Wilson coefficients in our $U(1)$ protected model (for $\alpha_2 \sim 1$).
The numerical value of the suppression scale is evaluated for two representative choices $f_{q_u^3}$,
and for  $g_{s*} = 6$,  $Y_{*} = 1$, $\delta = Y_*^2/16 \pi^2 (\Lambda R') \approx  Y_*^2/4 \pi^2 $.
In the last column we give the model independent lower bound on the suppression scale from ref. \cite{UTfit}.
}
\label{tab:all}
\end{table}
\renewcommand{\arraystretch}{1}

We now have everything we need to discuss flavor constraints on our model.
Inserting the experimentally measured values of the quark masses and mixing angles at the TeV scale (see Table 1 in \cite{CFW}), and making some assumptions concerning $g_{s*}$, $Y_*$ and $c_{q_u}^3$, we can find how much are the Wilson coefficients suppressed with respect to $1/M_G^2$.
Comparing it with the experimental lower bound we obtain a constraint on $M_G$, that is on the KK scale.
Sample results are given in Table \ref{tab:all}, where we took $g_{s*} = 6$, $Y_* = 1$.\footnote{In the standard RS flavor scenario one usually takes $Y_* \sim 3$, close to the perturbativity bounds, in order to maximally suppress the dangerous FCNC's.
In our case the flavor bounds can be satisfied with Yukawa couplings in a safely perturbative region. The loop induced kinetic mixings are then also naturally suppressed since in this case
$\delta = Y_*^2/16 \pi^2 (\Lambda R') \approx  Y_*^2/4 \pi^2 \ll 1$.
}
In the kaon sector, the LR operator (who is the most troublesome in the standard RS
flavor scenario with anarchic flavor in the down sector) comes out
very suppressed in our model and does not pose any problems. As for
the LL operator, it satisfies the experimental bounds even in the
standard RS with a 3 TeV KK gluon. In our case, the LL operator is
down by the additional loop factor $\delta^2$ and thus is always
well below the experimental limits. That contribution could be
sizable only if $\delta \sim 1$, that is for large IR brane kinetic
terms.

Quite unlike the standard RS flavor scenario, the strongest bounds
come from the D-meson sector. In Table \ref{tab:all} we present our
results for two different choices for the localization of the 3rd
generation doublet quark. In both cases we take $c_{q_u}^3 < 1/2$
corresponding to the IR localization, but  in the first case
($f_{q_u^3} = 0.3$) the profile is fairly flat, while in  the second
case ($f_{q_u^3} = 0.3$) it is sharply localized in IR. The
strongest bound always comes from the LR current, that is from the
coefficient $C_D^4$, but as long as $Y_* > 1$ a 3 TeV KK gluon is
allowed (a larger Yukawa coupling leads to relaxing this bound). In
the case with $f_{q_u^3} = 1$  another constraint comes from the LL
operator yielding a  $2 \tev$ bound on $M_G$. In any case, the
bounds are weaker than the ones from electroweak precision tests
that push the KK scale to around  $3 \tev$ anyways \cite{ADMS}. The
reason is that the flavor constraints  in the D meson sector are less
severe than analogous constraints in the kaon sector since $D^0 -
\bar D^0$ mixing is more challenging concerning both the
theoretical  prediction in the SM and the actual measurement (see
e.g. \cite{P}). Thus, imposing flavor symmetries in the down sector
and leaving an anarchic flavor violation in the up sector leads to
very mild flavor bounds on the scale of new physics.

Some more constraints on our model may be obtained by investigating $\Delta F = 1$ processes.   
Those generated at the loop level are beyond the scope of this paper, however there are  also tree-level $\Delta F = 1$ FCNC transitions mediated by the electrically neutral KK modes of $SU(2)_L \times SU(2)_R$ gauge bosons (the KK Z). 
The off-diagonal couplings of KK Z to the
left-handed quarks are analogous to those of the KK gluon:
$(g_{L}^{Z'})_{ij} \sim  (\pm 1/2) g_{w*} f_{q_u^i} f_{q_u^j}$, where
$g_{w*}$ is the $SU(2)$ bulk gauge coupling ($g_{w*} \sim 4$ in the absence of brane kinetic terms).
Electroweak symmetry breaking  mixes the KK Z with the SM Z boson via
the IR brane localized Higgs, with a mixing mass of order $g_{w*}m_Z v /2$ for both $SU(2)_L$ and  $SU(2)_R$ gauge bosons. 
This mixing induces the effective coupling  of the left-handed up quarks to the Z boson, for example  
$(g_{L}^{Z})_{ct} \sim  (g_{w*}^2/4) \lambda^2 f_{q_u^3}^2 m_Z v/M_{Z'}^2$, where $M_{Z'}$ is the mass of the lightest KK Z.
Note that for the down-type quarks $(g_{L}^{Z})_{bs}$ is much more suppressed because the $L$ and $R$ KK exchange approximately cancels in that case.  
The right-handed quarks, on the other hand, live in the $SU(2)_L \times SU(2)_R$ singlets so
that their effective couplings to the Z boson are also  suppressed.
To describe the effective $t_L Z c_L$ coupling, the low energy theory  contains the dimension six operator (${\tilde H} = i \sigma_2 H^*$)
\beq
\label{e.dupa2}
\mathcal{O}_{LL}^u = i\left[ {\overline Q}_3 {\tilde H} \right]
  \left[ \big( D\!\!\!\!\slash {\tilde H}^\dagger \big) Q_2 \right]
  - i\left[ {\overline Q}_3 \big( D\!\!\!\!\slash {\tilde H} \big) \right]
  \left[ {\tilde H}^\dagger Q_2 \right] + {\mathrm {h.c.}} 
\eeq 
with the coefficient 
\beq
C_{LL}^u \sim  {1 \over 2 M_{Z'}^{2}} g_{w*}^2 \lambda^2 f_{q_u^3}^2 
\sim  0.04 \left ( {g_{w*}  \over 4} \right )^2 \left (f_{q_u^3}  \over 1 \right )^2 \left ( {3 \tev \over M_{Z'}^{2}} \right )^2  \tev^{-2}
\eeq 
We took into account the fact the $L_\mu$ and $R_\mu$ have almost degenerate KK modes, and we assumed both the $L$ and $R$
coupling to be $g_{w*}$.
The coefficients of the analogous operators involving an up quark are more suppressed by RS-GIM. 
Now, at the loop level the operator in \eref{dupa2} feeds in to $B \to X_s \gamma$ and $B \to X_s l^+ l^-$ \footnote{We thank Uli Haisch for pointing this out to us.}, which leads to the bound $|C_{LL}^u| < 0.07 \tev^{-2}$ \cite{FLPPS}. 
This is satisfied for a 3 TeV KK scale, as  $f_{q_u^3}$ is not much larger than 1.

If $U(1)_d \times U(1)_q$ is   promoted to a local symmetry (as
expected from the AdS/CFT correspondence), there will be the
corresponding KK gauge bosons which couple to the SM quarks.\footnote{
However, one may as well use two discrete symmetries $Z_N\times Z_M$
with charges and symmetry breaking patterns identical to our
U(1)$\times$U(1) symmetries. In that case we would not have to
introduce U(1) gauge bosons. We thank Yuval Grossman for
this suggestion.} In case that the U(1) gauge bosons are present,
they will impose additional constraints on the parameters of the
model. First of all, we have to worry about the limits from direct
searches for new light neutral gauge bosons (aka Z'). The $U(1)_d$
KK modes do not pose a problem. As a consequence of the $[-+]$
boundary condition, they are localized in the IR and their masses
are at the KK scale which we assume to be at least $3 \tev$. On the
other hand, the lightest $U(1)_q$ gauge boson has an almost flat
profile (except near the IR brane) and its mass is suppressed with
respect to the KK scale (much as the W and Z masses in Higgsless
models~\cite{higgsless}): $M_q \sim (1/R') \sqrt{2/\log(R'/R)} \sim
.1 M_G$ which is around $300$ GeV when the lightest KK gluon is at 3
TeV. One way to evade the Tevatron limits is to assume that $U(1)_q$
does not couple to the SM leptons, another is to assume that the
$U(1)_q$ gauge coupling is small enough. In fact, the smallness of
the coupling is also required by the flavor bounds. Since the
$U(1)_q$ couplings to left-handed quarks is necessarily
non-universal, the rotation of the up-type quarks to the mass
eigenstate basis generates flavor non-diagonal couplings
\begin{equation}
(g_L^q)_{ij} \sim {g_{q*} \over
\log^{1/2}(R'/R)} |q_i - q_j| {f_{q_u^i} \over f_{q_u^j}} \qquad i < j
\end{equation}
where  $g_{q*}$ is the $U(1)_q$ bulk coupling, and the coupling of
the lightest KK mode is approximately $g_{q*}/\log^{1/2}(R'/R)$.
Unlike the KK gluon couplings, the flavor violating coupling to $u$
and $c$ is suppressed by only one power of the Cabibbo angle.
The RS-GIM mechanism is not acting here because this gauge boson is not IR brane localized, and the
SM fermions have family dependent couplings.
This fact together with  the
unbearable lightness of the $U(1)_q$ gauge boson implies that
$g_{q*}$ coupling must be much smaller than the strong bulk coupling
to pass the flavor bounds. More precisely, the tree-level exchange
of the $U(1)_q$ gauge boson contributes to the Wilson coefficient
$C_D^1$ as
\begin{equation}
C_D^1 \sim {g_{q*}^2 \over 2 \log (R'/R) M_q^2}
\lambda^2  \sim {g_{q*}^2 \over (12 \tev)^2} \left ( 3 \tev \over
M_G\right )^2
\end{equation}
Note, that a Fierz factor of $3$ in the denominator is missing
because a U(1) rather than SU(3) gauge boson is exchanged.
We can
see that we need $g_{q*} < 0.01$ to pass the experimental bounds listed in Table \ref{tab:all}
with a $3 \tev$ KK scale. 
Thus the $U(1)_q$ symmetry is required to
be global for all practical purposes. 

On the other hand, the bounds
on the $U(1)_d$ gauge coupling $g_{d*}$ are less severe, even though
the couplings of these KK modes do not have a full RS-GIM protection
either. The reason is that the $U(1)_d$ gauge bosons would be localized in the IR just like the KK gluon,
but the charges are now generation dependent so that the rotation to the
mass eigenstate basis leads to flavor violating couplings. The
largest contribution corresponds to the first  log suppressed  term
in (\ref{GBcoupling}), which could be ignored for the KK gluon since
it is generation independent in that case. After the rotation of the
left-handed up quarks, the leading non-diagonal couplings are $\sim
{g_{d*} \over  \log R'/R} L_{ij}^u |d_i \alpha_i^2 - d_j \alpha_j^2|$
and the contribution to the Wilson coefficient $C_D^1$ is given by
\begin{equation}
C_D^1 \sim {g_{d*}^2 \over 2 M_G^2} \frac{\lambda^{2}}{\log^2 R'/R} \alpha_2^4
\sim \frac{g_{d*}^2}{(700\ {\rm TeV})^2} \alpha_2^4 \left ( 3 \tev \over
M_G\right )^2.
\end{equation}
There could be an additional suppression by $\alpha_i$ because only quark component in
the 5D multiplet $\Psi_{q_d}$ is charged under $U(1)_d$,
and the left-handed up-type quarks could live dominantly in $\Psi_{q_u}$ if the $\alpha_i$ are small.
Without the additional $\alpha_i$ suppressions one would need $g_{d*}<1/2$ (somewhat weak), while with 
$\alpha_i$ suppression $g_{d*}$ may well be of the similar strength as the strong and electroweak bulk couplings.

Let us now discuss the experimental signatures.
The model we proposed predicts new physics contributions  to FCNC's in the
D-meson sector at the level close to the current experimental bound.
Improvement in both theory and experiment could lead to pinpointing new physics in the D-meson mixing and thus provide the first hint toward our model.
The model also predicts loop induced contributions to $\Delta F = 1$ processes in the down sector, for example to $B \to X_s \gamma$.    
On the other hand, a discovery of new sources of flavor and CP violation in the K- or B-meson mixing  would -- if not rule out
the model -- force it into an awkward region of parameter space with large brane kinetic terms.
At the LHC, the RS framework obviously predicts spectacular new
phenomena in abundance: KK modes of the SM fields, new gauge bosons
and exotic fermions required by the custodial symmetry,
see \cite{RSsig} for some recent papers.
Our flavor model does  not add many new states to that lot, though we may hope for a $\sim 3$ TeV $U(1)_d$ gauge boson with
non-universal couplings to the SM  quarks.
Some more hints from the  LHC may be revealed by FCNC processes
involving the top quark where the RS flavor models are known to
significantly enhance the SM predictions \cite{APStop}.
One sensitive probe is the decay  process  $t \to c Z$, which is
observable at the LHC at the level of $10^{-5}$ branching ratio.
However, in our model this  decay proceeds mostly via left-handed quarks,   
and there is the tension between obtaining a large enough coefficient $C_{LL}^u$ and the indirect constraint from $B \to X_s \gamma$ \cite{FLPPS}.   
This makes the prospects for a signal at the LHC challenging.

Finally, we comment on extending our idea to models of gauge-higgs
unification, where the Higgs boson is a pseudo-Goldstone boson whose
mass is protected by approximate global symmetries. For example, in
the model of ref. \cite{CDP} based on the $SU(3) \times SO(5) \times
U(1)$ gauge group in the bulk, the SM quark doublets also are
embedded in two different bulk multiplets, so that our U(1)
protection can be implemented. An important difference with respect
to our RS model is that in gauge-higgs unification the kinetic
mixing feeds the  flavor violation  into the left-handed down quark
sector even in the absence of brane kinetic terms \cite{CFW}.
Instead, the kinetic mixing originates from the same source as the Yukawa couplings (which is the boundary mass terms), therefore it is not loop suppressed as in our RS model. Nevertheless, in the limit of
unbroken electroweak symmetry there is no flavor violation in the
right-handed down sector  which suppresses the most dangerous
contributions to the LR four-fermion operators in the kaon sector.
The gauge-higgs model thus displays a similar phenomenology to the
model of ref. \cite{S}, with sizable flavor violation in both up and
down quark sectors.

In summary, we have presented a warped flavor model where most of
the flavor violation is moved into the up sector.
The key elements  are introducing two horizontal U(1) symmetries, and embedding the LH quark doublets into two separate bulk multiplets.
As a result,
all bulk mass parameters are aligned with the down-type Yukawa
coupling $\tilde{Y}_d$ and the extra mixing parameter $\theta$ (the
latter is needed due to the double embedding of the LH quarks). The
only source for flavor violation in the down sector are small IR
brane localized kinetic mixing terms. As a consequence, the leading
constraints arise from charm (D) physics (rather than K or B
physics), but these are also all satisfied with a KK mass scale of 3
TeV. If our U(1)'s are gauged, then these new gauge bosons will
introduce new sources of flavor violation, setting stringent bounds
on the possible sizes of these new gauge couplings.

{\it Acknowledgements}: We thank Yuval Grossman, Uli Haisch, and Matthias Neubert for useful discussions and comments and Zoltan Ligeti for sending us an updated version of \cite{FLPPS}.
C.C. thanks the KITP at UC Santa Barbara for its hospitality during the completion of this work.
The research of C.C. and A.W. is supported in part by the NSF grant PHY-0355005.
C.C. was supported in part by the NSF grant PHY05-51164 while at the KITP.
A.F. is partially supported by the European Community Contract MRTN-CT-2004-503369 for the years 2004--2008.


\end{document}